# A Collective Neurodynamic Approach to Survivable Virtual Network Embedding

Ashraf A. Shahin[1,2]

[1]College of Computer and Information Sciences, Al Imam Mohammad Ibn Saud Islamic University (IMSIU)
Riyadh, Kingdom of Saudi Arabia

[2]Department of Computer and Information Sciences, Institute of Statistical Studies & Research,
Cairo University, Cairo, Egypt

*Abstract*—Network virtualization has attracted significant amount of attention in the last few years as one of the key features of cloud computing. Network virtualization allows multiple virtual networks to share physical resources of single substrate network. However, sharing substrate network resources increases impact of single substrate resource failure. One of the commonly applied mechanisms to protect against such failures is provisioning redundant substrate resources for each virtual network to be used to recover affected virtual resources. However, redundant resources decreases cloud revenue by increasing virtual network embedding cost. In this paper, a collective neurodynamic approach has been proposed to reduce amount of provisioned redundant resources and reduce cost of embedding virtual networks. The proposed approach has been evaluated by using simulation and compared against some existing survivable virtual network embedding techniques.

*Keywords*—*Collective neurodynamics; integer linear programming; global optimization; network virtualization; survivable virtual network embedding*

## I. INTRODUCTION

Virtualization is one of the distinctive features of cloud computing. Virtualization increases utilization of substrate resources and increases revenue of cloud datacenters by allowing embedding multiple virtual networks in a single substrate network. However, mapping virtual resources to substrate resources is known to be NP-hard even without considering other cloud computing features such as scalability and survivability [1]-[3].

Although, sharing substrate resources among multiple virtual networks sustains cloud computing with many valuable benefits, it brings critical survivability issues. Single substrate resource failure can cause long service downtime and waste a lot of date from several virtual networks (VNs) [4]. Substrate resource failure becomes a part of everyday operation in today's Internet Service Provider (ISP) networks [5].

One of the most efficient protection approaches is provisioning redundant resources for each virtual network (VN). Redundant resources enable fast reallocating affected virtual resources after substrate resource failures. Nevertheless, redundant resources increase capacity of required virtual resources, which reduces revenue and reduces acceptance ratio of cloud datacenters.

In this paper, a collective neurodynamic optimization approach has been proposed to reduce amount of required redundant resources and to optimize virtual network embedding. To guarantee virtual network restorability after substrate node failure, the proposed approach enhances virtual network by adding one virtual node and set of virtual links. Virtual networks are enhanced by applying virtual network enhancing design proposed by Guo et al. in [1]. The problem has been formulated as Mixed-Integer Linear Programming and solved by applying neural network proposed by Xia in [6]. To guarantee survivability against substrate link failure, virtual links are embedded by applying multi-path link embedding approach proposed by Khan et al. in [7].

The problem of multi-path link embedding of enhanced virtual network has been formulated as Mixed-Integer Linear Programming and has been solved by using collective neurodynamic optimization approach, which combines the ability of social thinking in Particle Swarm Optimization with the local search capability of Neural Network.

Effectiveness of the proposed approach has been evaluated by comparing its performance with other approaches. Simulation results show that the proposed model reduces required redundant resources and increases revenue.

The rest of this paper is organized as follows. Section 2 describes the related work. Section 3 briefly describes the proposed model. Section 4 experimentally demonstrates the effectiveness of the proposed model. Finally, Section 6 concludes.

## II. RELATED WORK

Several survivable virtual network embedding (SVNE) approaches have been proposed in the last few years [1]-[4]. Guo et al. [1] have proposed survivable virtual network embedding approach. The proposed approach enhanced virtual network by adding additional virtual resources and redesigning virtual network with considering failure dependent protection technique, which provides backup substrate node for each substrate node failure scenario. Enhanced virtual network has been formulated using binary quadratic programming, and virtual network embedding has been formulated using mixed integer linear programming. Although, the proposed approach reduces amount of required substrate resources to design survivable virtual network, it increases number of required migrations after failures, which increases service down time.

A topology-aware remapping policy has been proposed by Xiao et al. [2] to deal with single substrate node failures. Based on network topology, a set of candidate backup substrate nodes has been defined for each substrate node and a set of candidate backup substrate links has been defined for each substrate link. In [8], Xiao et al. have extended the proposed policy in [2] to handle multiple nodes failures. However, the proposed policy uses all substrate nodes to accommodate incoming virtual networks. Therefore, when a substrate node failure happens, the proposed policy does not grantee that for each migrated virtual node there is a candidate backup substrate node with enough free resources to accommodate migrated virtual node.

Zhou et al. [3] have studied survivability of virtual networks against multiple physical link failures. They have formulated the problem of remapping virtual network with multiple physical link failures using mixed integer linear programming and have proposed an approach to find exact solution for the formulated problem. However, the proposed approach can deal only with small virtual networks.

Qiang et al. [9] have modeled the survivable virtual network embedding problem as an integer linear programming model and have used bee colony algorithm to find near optimal virtual network embedding solution. After substrate node failure, virtual nodes are migrated to normal node, which is specified by greed rules first, and virtual links are migrated to shortest substrate path. However, finding suitable substrate node for each migrated virtual node is a very complicated task. The reasons behind this complexity are not only connectivity and CPU constraints, but also lack of substrate resources. The proposed approach does not reserve a backup quota to be used after failures, which decreases the probability of finding enough and free substrate resources for recovering affected virtual resources and increases the probability of violating service level agreement.

To enhance virtual network survivability against single substrate link failure, Chen et al. [10] have proposed a linear programming model to formulate problem of allocating bandwidth for primary paths, backup paths, and shared backup paths. Performance of bandwidth allocation scheme has been improved by employing load-balancing strategy. After link failure, instead of migrating affected virtual links from failed substrate link, allocated bandwidths are reconfigured to cover affected virtual links.

Gu et al. [11], [13], [14] have proposed virtual network embedding scheme to guarantee recovery from single regional failure event. Before embedding each virtual network requests, working and recovery embeddings are specified to grantee that it is failure region disjoint. The problem has been formulated as a mixed integer linear programming problem and has been solved by proposing two heuristic solutions. The proposed solutions have improved resource efficiencies by considering mapping cost and load balancing during embedding process.

Mijumbi et al. [12] have proposed a distributed negotiation protocol to support virtual network survivability against physical link failures in multi-domain environments. The proposed protocol contains seven messages, which forms interactions between infrastructure providers and virtual network providers during backup and restoration for single physical link failures.

Meixner et al. [15] have proposed a probabilistic model to reduce the probability of virtual network disconnections and capacity loss due to single substrate link failure. The problem has been modeled as integer linear program model based on risk assessment to economically select sustainable backup substrate node.

Sun et al. [16] have modeled the problem of survivable virtual network embedding using mixed-integer linear programming (MILP) and proposed two algorithms: Lagrangian Relaxation based algorithm, and Decomposition based algorithm. While experimental shows that the proposed algorithms reduce computational complexity compared to MILP, they sustain the same embedding cost.

Rahman and Boutaba [5] have provided a mathematical formulation of the survivable virtual network embedding (SVNE) problem. To avoid mixed-integer programs complexity, Rahman and Boutaba have proposed hybrid policy heuristic, which proactively solves the problem of single substrate link failure with considering customer service level agreement constraints. Before any virtual network arrives, a set of candidate backup detours is calculated for each substrate link. After receiving new virtual network request, virtual nodes are embedded by applying two node-embedding algorithms: greedy node embedding heuristic that has been proposed by Zhu and Ammar in [17], and D-ViNE Algorithm, which has been proposed by Chowdhury and Rahman in [18]. To recover from substrate link failure, they have proposed hybrid and proactive backup detour algorithms. Performance of different combination of virtual node embedding algorithms and substrate link failure recovery algorithms have been evaluated by using ViNE-Yard simulator with different VN request topologies. Simulation results demonstrate that the proposed solutions outperform BLIND policy heuristic, which re-computes a new link embedding for each VN affected by the substrate link failure.

### III. Collective Neurodynamic Survivable VNE

Cloud datacenter receives users' requests as virtual networks (VNs). Each VN contains a set of virtual nodes (virtual machines) and a set of virtual links connect these nodes. Virtual network $VN$ is modeled as a weighted undirected graph $G_v = (N_v, L_v)$, where $N_v$ is the set of virtual nodes and $L_v$ is the set of virtual links. Virtual nodes and virtual links are weighted by the required CPU and bandwidth, respectively.

***Enhancing virtual network*** to improve survivability of virtual network $G_v = (N_v, L_v)$ with $n = |N_v|$ virtual nodes, an enhanced virtual network $G_e = (N_e, L_e)$ is constructed. Enhanced virtual network $G_e$ extends $G_v$ by adding one additional virtual node and set of virtual links. Resources of $G_e$ are specified to guarantee that there are enough resources to reallocate $G_v$ after any substrate link or single node failures.

Each reallocation is represented by using three matrices $A$, $B$, and $C$. $A$ is $(n + 1)$ matrix, where $a_i = j$ means that virtual node $i$ is allocated on node j. $B$ is $(n + 1) \times (n + 1)$ matrix, where $b_{ij}$ refers to the required bandwidth for the virtual link that connects virtual nodes $i$ and $j$. $C$ is $(n+1)$ matrix, where $c_i = y$ implies that virtual node $i$ requires CPU with capacity $y$.

Virtual network enhancing process is initialized by allocating virtual nodes from virtual network (which contains $n$ virtual nodes) to the first $n$ nodes in the enhanced virtual network. The initial allocation matrix $A^0$ becomes as following

$$A^0 = (1, 2, .., n, 0)$$

$a_{n+1}^0 = 0$ refers to that node $n+1$ in the enhanced virtual network is empty and is not used in this case.

To recovery from $k^{th}$ substrate node failure, re-embedding matrix $A^k$ can be generated by permuting matrix $A^0$. Matrix $A^k$ can be represented by permutation matrix $X^k$, which is an orthogonal matrix of size $(n+1) \times (n+1)$ such that

$$x_{ij}^k = \begin{cases} 1 & \text{if } a_i^k = j, \\ 0 & \text{otherwise.} \end{cases}$$

Thus, re-embedding matrices is calculated as

$$A^k = A^0 X^k$$
$$C^k = C^0 X^k$$
$$B^k = (X^k)^T B^0 X^k$$

Therefore, the problem of enhancing virtual network is formulated as following:

Min $\left( \sum_{i=1}^{n+1} c_i^e + \alpha \sum_{i=1}^{n+1} \sum_{j=1}^{n+1} b_{ij}^e \right)$ (1)

Subject to

$c_i^e = \max_k c_i^k, \forall k \in \{1, .., n\}, i \in \{1, .., n+1\}$ (2)

$b_{ij}^e = \max_k b_{ij}^k, \forall k \in \{1, .., n\}, i,j \in \{1, .., n+1\}$ (3)

$c_i^k = \sum_{j=1}^{n+1} c_j^0 x_{ji}^k, \forall k \in \{1, .., n\}, i \in \{1, .., n+1\}$ (4)

$b_{ij}^k = \sum_{m=1}^{n+1} \sum_{l=1}^{n+1} b_{lm}^0 x_{li}^k x_{mj}^k, \forall k \in \{1, .., n\}, i,j \in \{1, .., n+1\}$ (5)

$\sum_{j=1}^{n+1} x_{ij}^k = 1, \forall k \in \{1, .., n\}, i \in \{1, .., n+1\}$ (6)

$\sum_{i=1}^{n+1} x_{ij}^k = 1, \forall k \in \{1, .., n\}, j \in \{1, .., n+1\}$ (7)

$x_{ik}^k = 1, \forall k \in \{1, .., n\}, i = n+1$ (8)

$x_{ij}^k \in \{0,1\}, \forall k \in \{1, .., n\}, i,j \in \{1, .., n+1\}$ (9)

Where, α is the weight coefficient to represent importance of bandwidth and CPU resources. Constraints (2) - (5) ensure that there are sufficient resources to re-allocate virtual network after different failures. Constraint (6) reveals that each virtual node is allocated to only one substrate node and constraint (7) makes certain that each substrate node contains only one virtual node from the same virtual network. Constraint (8) guarantees that for each $k^{th}$ substrate node failure there is a permutation matrix $X^k$ to generate re-embedding matrix $A^k$ from the initial matrix $A^0$.

This problem is nonconvex due to bilinear constraints (5) and it can be linearized by replacing the quadratic terms $x_{li}^k x_{mj}^k$ by five-dimensional array $Y = (y_{limjk})$

$$y_{limjk} = x_{li}^k x_{mj}^k, \forall k \in \{1, .., n\}, l,i,m,j \in \{1, .., n+1\}$$

After replacing the zero-one integrality constraint (9) with non-negativity constraint, the problem becomes equivalent to the following linear programming problem:

Min $P(Z) = \left( \sum_{i=1}^{n+1} c_i^e + \alpha \sum_{i=1}^{n+1} \sum_{j=1}^{n+1} b_{ij}^e \right)$ (10)

Subject to

$\sum_{i=1}^{n+1} c_i^0 x_{ij}^k - c_j^e \leq 0, \forall k \in \{1, .., n\}, j \in \{1, .., n+1\}$ (11)

$\sum_{j=1}^{n+1} \sum_{i=1}^{n+1} b_{ij}^0 y_{limjk} - b_{lm}^e \leq 0, \forall k \in \{1, .., n\}, l,m \in \{1, .., n+1\}$ (12)

$\sum_{l=1}^{n+1} \sum_{i=1}^{n+1} \sum_{m=1}^{n+1} \sum_{j=1}^{n+1} y_{limjk} = (n+1)^2, \forall k \in \{1, .., n\}$ (13)

$2y_{limjk} - x_{li}^k - x_{mj}^k \leq 0, \forall k \in \{1, .., n\}, l,i,m,j \in \{1, .., n+1\}$ (14)

$y_{limjk} \geq 0, \forall k \in \{1, .., n\}, i,j,l,m \in \{1, .., n+1\}$ (15)

$\sum_{j=1}^{n+1} x_{ij}^k = 1, \forall k \in \{1, .., n\}, i \in \{1, .., n+1\}$ (16)

$\sum_{i=1}^{n+1} x_{ij}^k = 1, \forall k \in \{1, .., n\}, j \in \{1, .., n+1\}$ (17)

$x_{ik}^k = 1, \forall k \in \{1, .., n\}, i = n+1$ (18)

$x_{ij}^k \geq 0, \forall k \in \{1, .., n\}, i,j \in \{1, .., n+1\}$ (19)

Z is defined as $Z = (C^{e\prime}, B^{e\prime}, Y', X')^T$, where $C^{e\prime}$ and $B^{e\prime}$ are vectors represent required CPU and bandwidth for the enhanced virtual network. $Y'$ and $X'$ are vectors represent all variables in the five-dimensional array Y and in the three-dimensional array X, which contains all permutations.

The primal problem (10)-(19) is liner programming and it can be written in the following general matrix form:

Minimize $\quad d^T z$ (20)

Subject to

$M_{11}z_1 + M_{12}z_2 \geq r_1$ (21)

$M_{21}z_1 + M_{22}z_2 = r_2$ (22)

$z_1 \geq 0$ (23)

where $z = \begin{pmatrix} z_1 \\ z_2 \end{pmatrix} \in \mathbb{R}^{q_1}$, $d = \begin{pmatrix} d_1 \\ d_2 \end{pmatrix} \in \mathbb{R}^{q_1}$,

$$r = \begin{pmatrix} r_1 \\ r_2 \end{pmatrix} \in \mathbb{R}^{q_2},$$

$$q_1 = n + n^2 + n.(n+1)^2 + n.(n+1)^4,$$

$$q_2 = n^5 + 2n^4 + 3n^3 + 8n^2 + 7n$$

From dual theory [6], dual problem for the previous primal problem is:

Maximize $\quad r^T \xi$ (24)

Subject to

$$M_{11}\xi_1 + M_{21}\xi_2 \leq d_1 \quad (25)$$

$$M_{12}\xi_1 + M_{22}\xi_2 = d_2 \quad (26)$$

$$\xi_1 \geq 0 \quad (27)$$

Where $\xi = (\lambda, \mu)^T$, and $\xi = \begin{pmatrix} \xi_1 \\ \xi_2 \end{pmatrix} \in \mathbb{R}^{q_2}$

To solve the primal problem (20)-(23) and dual problem (24)-(27), Xia [6] has proposed neural network with the following differential equation to drive its state vector $u = (z, \xi)^T$.

$$\frac{du}{dt} = -\nabla E(u) \quad (28)$$

Where, $\nabla E(u)$ is the gradient of the energy function $E(u)$, which is defined as following:

$$E(z, \xi) = \frac{1}{2}(d^T z - r^T \xi)^2 + \frac{1}{2} z_1^T(z_1 - |z_1|)$$

$$+ \frac{1}{2}\xi_1^T(\xi_1 - |\xi_1|)$$

$$+ \frac{1}{2}\|M_2 z - r_2\|_2^2 + \frac{1}{2}\|M_4 \xi - d_2\|_2^2$$

$$+ \frac{1}{2}[M_1 z - r_1]^T [(M_1 z - r_1) - |M_1 z - r_1|]$$

$$+ \frac{1}{2}[d_1 - M_3 \xi]^T [(d_1 - M_3 \xi) - |d_1 - M_3 \xi|], \quad (29)$$

Where, $M_1 = (M_{11}, M_{12})$, $M_2 = (M_{21}, M_{22})$,

$$M_3 = (M_{11}^T, M_{21}^T), \text{ and } M_4 = (M_{12}^T, M_{22}^T)$$

To apply the previous neural network to the primal problem (10)-(19), constraints are switched to penalties by using Lagrange and Karush-Kuhn-Tucker multipliers as follows:

$$L(z, \lambda, \mu) = \left(\sum_{i=1}^{n+1} c_i^e + \alpha \sum_{i=1}^{n+1}\sum_{j=1}^{n+1} b_{ij}^e\right)$$

$$+ \sum_{k=1}^{n} \lambda_k \cdot \left(\sum_{l=1}^{n+1}\sum_{i=1}^{n+1}\sum_{m=1}^{n+1}\sum_{j=1}^{n+1} y_{limjk} - (n+1)^2\right)$$

$$+ \sum_{i=1}^{n+1}\sum_{k=1}^{n} \lambda_{k+n.i} \cdot \left(\sum_{j=1}^{n+1} x_{ij}^k - 1\right)$$

$$+ \sum_{j=1}^{n+1}\sum_{k=1}^{n} \lambda_{k+n.j+n.(n+1)} \cdot \left(\sum_{i=1}^{n+1} x_{ij}^k - 1\right)$$

$$+ \sum_{i=1}^{n+1}\sum_{k=1}^{n} \lambda_{k+n.i+2.n.(n+1)} \cdot \left(x_{ik}^k - 1\right)$$

$$+ \sum_{j=1}^{n+1}\sum_{k=1}^{n} \mu_{q_{kj}} \cdot \left(\sum_{i=1}^{n+1} c_i^0 x_{ij}^k - c_j^e\right)$$

$$+ \sum_{l=1}^{n+1}\sum_{m=1}^{n+1}\sum_{k=1}^{n} \mu_{q_{kml}} \cdot \left(\sum_{j=1}^{n+1}\sum_{i=1}^{n+1} b_{ij}^0 y_{limjk} - b_{lm}^e\right)$$

$$+ \sum_{l=1}^{n+1}\sum_{m=1}^{n+1}\sum_{i=1}^{n+1}\sum_{j=1}^{n+1}\sum_{k=1}^{n} \mu_{q_{kjiml}} \cdot \left(2 y_{limjk} - x_{li}^k - x_{mj}^k\right), \quad (30)$$

$$\mu_i \geq 0, \forall i \in \{1, \ldots, n^5 + 2n^4 + 3n^3 + 5n^2 + 3n\}$$

Where, $\lambda$ and $\mu$ are vectors of Lagrange and Karush-Kuhn-Tucker multipliers, respectively.

$$\lambda = (\lambda_1, \ldots, \lambda_i), i \in \{1, \ldots, 3n^2 + 4n\}$$

$$\mu = (\mu_1, \ldots, \mu_i), i \in \{1, \ldots, n^5 + 2n^4 + 3n^3 + 5n^2 + 3n\}$$

$q_{kj}, q_{kml},$ and $q_{kjiml}$ are defined as following:

$$q_{kj} = k + n.(j-1), \forall k \in \{1, \ldots, n\}, j \in \{1, \ldots, n+1\}$$

$$q_{kml} = k + n.(m+n) + n.(n+1).(l-1),$$
$$\forall k \in \{1, \ldots, n\}, m, l \in \{1, \ldots, n+1\}$$

$$q_{kjiml} = k + n.j + n^2 + n.(n+1).(i + n.m + n^2.l - n^2),$$
$$\forall k \in \{1, \ldots, n\}, i, j, l, m \in \{1, \ldots, n+1\}$$

To find the dual function, which is infimum value of the lagrangian function $L(z, \lambda, \mu)$, the derivative of the function $L(z, \lambda, \mu)$ with respect to $z$ must be zero. Thus, the dual problem is formulated as follows:

Maximize

$$D(\lambda, \mu) = -\sum_{k=1}^{n}(n+1)^2.\lambda_k - \sum_{i=1}^{n+1}\sum_{k=1}^{n} \lambda_{k+n.i}$$
$$- \sum_{j=1}^{n+1}\sum_{k=1}^{n} \lambda_{k+n.j+n.(n+1)} -$$
$$\sum_{i=1}^{n+1}\sum_{k=1}^{n} \lambda_{k+n.i+2.n.(n+1)} \quad (31)$$

Subject to

$$1 - \sum_{k=1}^{n} \mu_{k+n.(j-1)} = 0, \forall j \in \{1, \ldots, n+1\} \quad (32)$$

$$\alpha - \sum_{k=1}^{n} \mu_{k+n.(j+n)+n.(n+1).(i-1)} = 0, \forall i, j \in \{1, \ldots, n+1\} \quad (33)$$

$$\lambda_k + b_{ij}^0 \cdot \mu_{q_{kml}} + 2 \mu_{q_{kjiml}} = 0, \forall i, j, l, m \in \{1, \ldots, n+1\}, k \in \{1, \ldots, n\} \quad (34)$$

$$\lambda_{k+n.i} + \lambda_{k+n.k+n.(n+1)} + \lambda_{k+n.i+2.n.(n+1)} +$$
$$c_i^0 \cdot \mu_{k+n.(k-1)} - \sum_{l=1}^{n+1}\sum_{m=1}^{n+1} \mu_{q'_{kmil}} = 0, \quad (35)$$

$$q'_{kmil} = k + n.k + n^2$$
$$+ n.(n+1).(m + n.i + n^2.l - n^2),$$
$$\forall i, m, l \in \{1, \ldots, n+1\}, k \in \{1, \ldots, n\}$$

$$\lambda_{k+n.i} + \lambda_{k+n.j+n.(n+1)} + c_i^0 \cdot \mu_{k+n.(j-1)} -$$
$$\sum_{l=1}^{n+1}\sum_{m=1}^{n+1} \mu_{q_{kjmil}} = 0, \quad (36)$$

$$q_{kjmil} = k + n.j + n^2$$
$$+ n.(n+1).(m + n.i + n^2.l - n^2),$$
$$\forall i, j, m, l \in \{1, \ldots, n+1\}, k \in \{1, \ldots, n\}, j \neq k$$

$$\mu_i \geq 0, \forall i \in \{1, \ldots, n^5 + 2n^4 + 3n^3 + 5n^2 + 3n\} \quad (37)$$

The time derivative of a state variable is calculated as partial derivative of the energy function in (28). Thus, the dynamic equation of the neural network is defined by the following differential equations:

$$\frac{dc_i^e}{dt} = -\beta \{ \sum_{m=1}^{n+1} c_m^e + \alpha \sum_{m=1}^{n+1} \sum_{j=1}^{n+1} b_{m\,j}^e +$$
$$\sum_{k=1}^{n} (n+1)^2 \cdot \lambda_k + \sum_{m=1}^{n+1} \sum_{k=1}^{n} \lambda_{k+n.m} +$$
$$\sum_{j=1}^{n+1} \sum_{k=1}^{n} \lambda_{k+n.j+n.(n+1)} +$$
$$\sum_{m=1}^{n+1} \sum_{k=1}^{n} \lambda_{k+n.m+2.n.(n+1)} + \sum_{k=1}^{n} (c_i^e - \sum_{j=1}^{n+1} c_j^0 x_{ji}^k - |c_i^e - \sum_{j=1}^{n+1} c_j^0 x_{ji}^k|) \}, \forall i \in$$
$$\{1,..,n+1\} \qquad (38)$$

$$\frac{db_{ij}^e}{dt} = -\beta \{ \alpha \cdot ( \sum_{l=1}^{n+1} c_l^e + \alpha \sum_{l=1}^{n+1} \sum_{m=1}^{n+1} b_{l\,m}^e +$$
$$\sum_{k=1}^{n} (n+1)^2 \cdot \lambda_k + \sum_{l=1}^{n+1} \sum_{k=1}^{n} \lambda_{k+n.l} +$$
$$\sum_{l=1}^{n+1} \sum_{k=1}^{n} \lambda_{k+n.l+n.(n+1)} +$$
$$\sum_{l=1}^{n+1} \sum_{k=1}^{n} \lambda_{k+n.l+2.n.(n+1)} ) + \sum_{k=1}^{n} (b_{ij}^e -$$
$$\sum_{l=1}^{n+1} \sum_{m=1}^{n+1} b_{ml}^0 y_{imjlk} -$$
$$|b_{ij}^e - \sum_{l=1}^{n+1} \sum_{m=1}^{n+1} b_{ml}^0 y_{imjlk}|) \}, \forall i,j \in \{1,..,n+1\} \qquad (39)$$

$$\frac{dy_{ijlmk}}{dt} =$$
$$-\beta \{ y_{ijlmk} - |y_{ijlmk}| +$$
$$\sum_{p=1}^{n+1} \sum_{q=1}^{n+1} \sum_{r=1}^{n+1} \sum_{s=1}^{n+1} y_{pqrsk} - (n+1)^2 -$$
$$b_{jm}^0 \cdot b_{il}^e + b_{jm}^0 \cdot \sum_{q=1}^{n+1} \sum_{p=1}^{n+1} b_{pq}^0 y_{iplqk} -$$
$$b_{jm}^0 \cdot |b_{il}^e - \sum_{q=1}^{n+1} \sum_{p=1}^{n+1} b_{pq}^0 y_{iplqk}| - 2 x_{ij}^k - 2 x_{lm}^k +$$
$$4 y_{ijlmk} + 2 |x_{ij}^k + x_{lm}^k - 2 y_{ijlmk}| \}, \forall i,j,l,m \in$$
$$\{1,..,n+1\}, k \in \{1,..,n\} \qquad (40)$$

$$\frac{dx_{ij}^k}{dt} = -\beta \{ x_{ij}^k - |x_{ij}^k| + \sum_{p=1}^{n+1} (x_{ip}^k + x_{pi}^k) - 2 -$$
$$c_p^0 \cdot c_j^e + c_p^0 \cdot \sum_{p=1}^{n+1} c_p^0 x_{pj}^k + c_p^0 \cdot |c_j^e - \sum_{p=1}^{n+1} c_p^0 x_{pj}^k| +$$
$$\sum_{q=1}^{n+1} \sum_{m=1}^{n+1} (x_{ij}^k + x_{mq}^k - 2 y_{ijmqk}) -$$
$$\sum_{q=1}^{n+1} \sum_{m=1}^{n+1} |x_{ij}^k + x_{mq}^k - 2 y_{ijmqk}| +$$
$$\sum_{l=1}^{n+1} \sum_{p=1}^{n+1} (x_{lp}^k + x_{ij}^k - 2 y_{lpijk}) -$$
$$\sum_{l=1}^{n+1} \sum_{p=1}^{n+1} |x_{lp}^k + x_{ij}^k - 2 y_{lpijk}| - 2 x_{ij}^k + 2 y_{ijijk} +$$
$$|2 x_{ij}^k - 2 y_{ijijk}| \}, \forall\ i,j \in \{1,..,n+1\}, k \in$$
$$\{1,..,n\}, such\ that\ (i \neq n+1)\ or\ (j \neq$$
$$k) \qquad (41)$$

$$\frac{dx_{(n+1)k}^k}{dt} = -\beta \{ x_{(n+1)k}^k - |x_{(n+1)k}^k| + \sum_{p=1}^{n+1} (x_{(n+1)p}^k + x_{p(n+1)}^k) - 2 - c_p^0 \cdot c_k^e + c_p^0 \cdot \sum_{p=1}^{n+1} c_p^0 x_{pk}^k +$$
$$c_p^0 \cdot |c_k^e - \sum_{p=1}^{n+1} c_p^0 x_{pk}^k| + \sum_{q=1}^{n+1} \sum_{m=1}^{n+1} (x_{(n+1)k}^k +$$
$$x_{mq}^k - 2 y_{(n+1)kmqk}) - \sum_{q=1}^{n+1} \sum_{m=1}^{n+1} |x_{(n+1)k}^k +$$
$$x_{mq}^k - 2 y_{(n+1)kmqk}| + \sum_{l=1}^{n+1} \sum_{p=1}^{n+1} (x_{lp}^k + x_{(n+1)k}^k -$$
$$2 y_{lp(n+1)kk}) - \sum_{l=1}^{n+1} \sum_{p=1}^{n+1} |x_{lp}^k + x_{(n+1)k}^k -$$
$$2 y_{lp(n+1)kk}| + x_{(n+1)k}^k - 1 \}, \forall k \in$$
$$\{1,..,n\} \qquad (42)$$

$$\frac{d\lambda_i}{dt} = -\beta \{ \sum_{s=1}^{n+1} \sum_{j=1}^{n+1} \sum_{l=1}^{n+1} \sum_{m=1}^{n+1} (\lambda_i + b_{sj}^0 \cdot \mu_{q_{iml}} +$$
$$2 \mu_{q_{ijsml}}) +$$
$$(n+1)^2 \cdot (\sum_{s=1}^{n+1} c_s^e + \alpha \sum_{s=1}^{n+1} \sum_{j=1}^{n+1} b_{s\,j}^e +$$
$$\sum_{k=1}^{n} (n+1)^2 \cdot \lambda_k + \sum_{s=1}^{n+1} \sum_{k=1}^{n} \lambda_{k+n.s} +$$
$$\sum_{j=1}^{n+1} \sum_{k=1}^{n} \lambda_{k+n.j+n.(n+1)} +$$
$$\sum_{s=1}^{n+1} \sum_{k=1}^{n} \lambda_{k+n.s+2.n.(n+1)} ) \} \forall i \in$$
$$\{1,..,n\} \qquad (43)$$

$$\frac{d\lambda_{k+n.s}}{dt} = -\beta \{ \sum_{j \in \{1,..,n+1\} - \{k\}} (\lambda_{k+n.s} + \lambda_{k+n.j+n.(n+1)} +$$
$$c_s^0 \cdot \mu_{k+n.(j-1)} - \sum_{l=1}^{n+1} \sum_{m=1}^{n+1} \mu_{q_{kjmsl}}) + \lambda_{k+n.s} +$$
$$\lambda_{k+n.k+n.(n+1)} + \lambda_{k+n.s+2.n.(n+1)} + c_s^0 \cdot \mu_{k+n.(k-1)} -$$
$$\sum_{l=1}^{n+1} \sum_{m=1}^{n+1} \mu_{q'_{kmsl}} + \sum_{i=1}^{n+1} c_i^e + \alpha \sum_{i=1}^{n+1} \sum_{j=1}^{n+1} b_{i\,j}^e +$$
$$\sum_{r=1}^{n} (n+1)^2 \cdot \lambda_r + \sum_{i=1}^{n+1} \sum_{r=1}^{n} \lambda_{r+n.i} +$$
$$\sum_{j=1}^{n+1} \sum_{r=1}^{n} \lambda_{r+n.j+n.(n+1)} +$$
$$\sum_{i=1}^{n+1} \sum_{r=1}^{n} \lambda_{r+n.i+2.n.(n+1)} \}, \forall k \in \{1,..,n\}, s \in$$
$$\{1,..,n+1\} \qquad (44)$$

$$\frac{d\lambda_{k+n.j+n.(n+1)}}{dt} = -\beta \{ \sum_{s=1}^{n+1} (\lambda_{k+n.s} + \lambda_{k+n.j+n.(n+1)} +$$
$$c_s^0 \cdot \mu_{k+n.(j-1)} - \sum_{l=1}^{n+1} \sum_{m=1}^{n+1} \mu_{q_{kjmsl}}) + \sum_{i=1}^{n+1} c_i^e +$$
$$\alpha \sum_{i=1}^{n+1} \sum_{r=1}^{n+1} b_{i\,r}^e + \sum_{s=1}^{n} (n+1)^2 \cdot \lambda_s +$$
$$\sum_{i=1}^{n+1} \sum_{s=1}^{n} \lambda_{s+n.i} + \sum_{r=1}^{n+1} \sum_{s=1}^{n} \lambda_{s+n.r+n.(n+1)} +$$
$$\sum_{i=1}^{n+1} \sum_{s=1}^{n} \lambda_{s+n.i+2.n.(n+1)} \} \forall k \in \{1,..,n\}, j \in$$
$$\{1,..,n+1\}, k \neq j \qquad (45)$$

$$\frac{d\lambda_{k+n.k+n.(n+1)}}{dt} = -\beta \{ \sum_{s=1}^{n+1} (\lambda_{k+n.s} + \lambda_{k+n.k+n.(n+1)} +$$
$$c_s^0 \cdot \mu_{k+n.(k-1)} - \sum_{l=1}^{n+1} \sum_{m=1}^{n+1} \mu_{q_{kkmsl}}) +$$
$$\sum_{s=1}^{n+1} (\lambda_{k+n.s} + \lambda_{k+n.k+n.(n+1)} + \lambda_{k+n.s+2.n.(n+1)} +$$
$$c_s^0 \cdot \mu_{k+n.(k-1)} - \sum_{l=1}^{n+1} \sum_{m=1}^{n+1} \mu_{q'_{kmsl}}) + \sum_{i=1}^{n+1} c_i^e +$$
$$\alpha \sum_{i=1}^{n+1} \sum_{r=1}^{n+1} b_{i\,r}^e + \sum_{s=1}^{n} (n+1)^2 \cdot \lambda_s +$$
$$\sum_{i=1}^{n+1} \sum_{s=1}^{n} \lambda_{s+n.i} + \sum_{r=1}^{n+1} \sum_{s=1}^{n} \lambda_{s+n.r+n.(n+1)} +$$
$$\sum_{i=1}^{n+1} \sum_{s=1}^{n} \lambda_{s+n.i+2.n.(n+1)} \}, \forall k \in$$
$$\{1,..,n\} \qquad (46)$$

$$\frac{d\lambda_{k+n.j+2.n.(n+1)}}{dt} = -\beta \{ \sum_{i=1}^{n+1} c_i^e + \alpha \sum_{i=1}^{n+1} \sum_{r=1}^{n+1} b_{i\,r}^e +$$
$$\sum_{s=1}^{n} (n+1)^2 \cdot \lambda_s + \sum_{i=1}^{n+1} \sum_{k=1}^{n} \lambda_{s+n.i} +$$
$$\sum_{r=1}^{n+1} \sum_{s=1}^{n} \lambda_{s+n.r+n.(n+1)} +$$
$$\sum_{i=1}^{n+1} \sum_{s=1}^{n} \lambda_{s+n.i+2.n.(n+1)} + \lambda_{k+n.j} +$$
$$\lambda_{k+n.k+n.(n+1)} + \lambda_{k+n.j+2.n.(n+1)} + c_j^0 \cdot \mu_{k+n.(k-1)} -$$
$$\sum_{l=1}^{n+1} \sum_{m=1}^{n+1} \mu_{q'_{kmj}} \}, \forall k \in \{1,..,n\}, j \in \{1,..,n+1\} \qquad (47)$$

$$\frac{d\mu_{k+n.(j-1)}}{dt} = -\beta \Big\{ \mu_{k+n.(j-1)} - |\mu_{k+n.(j-1)}| + \sum_{s=1}^{n} \mu_{s+n.(j-1)} - 1 + \sum_{i=1}^{n+1} \big(\lambda_{k+n.i} + \lambda_{k+n.j+n.(n+1)} + c_i^0 . \mu_{k+n.(j-1)} - \sum_{l=1}^{n+1}\sum_{m=1}^{n+1} \mu_{q_{kjmil}}\big) \Big\}, \forall\, j \in \{1,..,n+1\}, k \in \{1,..,n\}, j \neq k \quad (48)$$

$$\frac{d\mu_{k+n.(k-1)}}{dt} = -\beta \Big\{ \mu_{k+n.(k-1)} - |\mu_{k+n.(k-1)}| + \sum_{s=1}^{n} \mu_{s+n.(k-1)} - 1 + \sum_{s=1}^{n+1} \big(\lambda_{k+n.s} + \lambda_{k+n.k+n.(n+1)} + \lambda_{k+n.s+2.n.(n+1)} + c_s^0 . \mu_{k+n.(k-1)} - \sum_{l=1}^{n+1}\sum_{m=1}^{n+1} \mu_{q'_{kmsl}}\big) \Big\}, q'_{kmsl} = k + n.k + n^2 + n.(n+1).(m+n.s+n^2.l - n^2), \forall\, k \in \{1,..,n\} \quad (49)$$

$$\frac{d\mu_{q_{kji}}}{dt} = -\beta \Big\{ \mu_{q_{kji}} - |\mu_{q_{kji}}| + \sum_{s=1}^{n} \mu_{q_{sji}} - \alpha + \sum_{s=1}^{n+1}\sum_{r=1}^{n+1} \big(\lambda_k + b_{sr}^0 . \mu_{q_{kji}} + 2\mu_{q_{krsji}}\big) \Big\}, q_{kji} = k + n.(j+n) + n.(n+1).(i-1), \forall\, i,j \in \{1,..,n+1\}, k \in \{1,..,n\}$$

$$\frac{d\mu_{q_{kjmil}}}{dt} = -\beta \Big\{ \mu_{q_{kjmil}} - |\mu_{q_{kjmil}}| + \lambda_{k+n.i} + \lambda_{k+n.j+n.(n+1)} + c_i^0.\mu_{k+n.(j-1)} - \sum_{s=1}^{n+1}\sum_{r=1}^{n+1} \mu_{q_{kjris}} + \lambda_k + b_{mj}^0 . \mu_{q_{kil}} + 2\mu_{q_{kjmil}} \Big\}, q_{kjmil} = k + n.j + n^2 + n.(n+1).(m+n.i+n^2.l - n^2), \forall\, i,j,m,l \in \{1,..,n+1\}, k \in \{1,..,n\}, j \neq k \quad (50)$$

$$\frac{d\mu_{q'_{kmil}}}{dt} = -\beta \Big\{ \mu_{q'_{kmil}} - |\mu_{q'_{kmil}}| + \lambda_{k+n.i} + \lambda_{k+n.k+n.(n+1)} + \lambda_{k+n.i+2.n.(n+1)} + c_i^0.\mu_{k+n.(k-1)} - \sum_{s=1}^{n+1}\sum_{r=1}^{n+1} \mu_{q'_{kris}} + \lambda_k + b_{mk}^0 . \mu_{q_{kil}} + 2\mu_{q'_{kmil}} \Big\}, q'_{kmil} = k + n.k + n^2 + n.(n+1).(m+n.i+n^2.l-n^2), q_{kil} = k + n.(i+n) + n.(n+1).(l-1), \forall\, i,m,l \in \{1,..,n+1\}, k \in \{1,..,n\} \quad (51)$$

Where, $\beta$ is a nonnegative parameter that scales the convergence rate of the neural network.

***Survivable virtual network embedding*** after describing neural network that has been used for enhancing virtual network, this subsection explains the collective neurodynamic approach that has been employed for finding optimal multi-path link embedding solution.

Multiple neurodynamic models have been exploited to enhance candidate virtual network embedding solution. First, set of neurodynamic models are initialized and distributed in the substrate network. Second, each model improves its local optimal solution by using its dynamic equation. After all neurodynamic models converge to its local optimal, Particle Swarm Optimization (PSO) technique is employed to exchange information between neurodynamic models. Third, each model adjusts its state with considering its own best solution as well as the best solution found so far. The previous two steps are repeated until a termination criterion is reached. The position and velocity of each neurodynamic model can be expressed using the following equations [19]:

$$V_i(t+1) = wV_i(t) + c_1 r_1 \big(pBest_i(t) - X_i(t)\big) + c_2 r_2 \big(gBest(t) - X_i(t)\big) \quad (52)$$

$$X_i(t+1) = X_i(t) + V_i(t+1) \quad (53)$$

Where, $pBest_i(t)$ is the local optimal solution of the neurodynamic model i at time t, $gBest(t)$ is the global optimal solution at time t, $r_1$ and $r_2$ are two random numbers between 0 and 1. The constants $c_1$, and $c_2$ are specified to control influence of $pBest_i(t)$ and $gBest(t)$ on the search process. The constant w is called inertia weight, which controls effect of the previous velocity on the new one.

In the remaining of this subsection, the problem of multi-path link embedding of enhanced virtual network has been formulated as quadratic integer program and transformed into mixed integer linear program. Finally, dynamic equation of the neurodynamic model has been explained.

In multi-path link embedding [7], each virtual link is divided into $\eta > 1$, $\eta \in \mathbb{Z}$ virtual sub-links, which connect the same virtual nodes as the original virtual link. Bandwidth of each virtual sub-link is equal to $1/(\eta - 1)$ of the original virtual link bandwidth. Consequently, there is only one extra sub-link is added for survivability against substrate link failure. The problem of embedding enhanced virtual network is formulated as quadratic integer program:

$$Min \; \sum_{i=1}^{n}\sum_{j=1}^{n}\sum_{k=1}^{m}\sum_{l=1}^{m} Length(p_{kl}) x_{ik} x_{jl} b_{ij} \quad (54)$$

$$(c_i^v - c_k^s) x_{ik} \leq 0, \forall\, i \in \{1,..,n\}, k \in \{1,..,m\} \quad (55)$$

$$\Big(\frac{1}{\eta-1} b_{ij} - Band(p_{kl})\Big) x_{ik} x_{jl} \leq 0, \forall\, i,j \in \{1,..,n\}, k,l \in \{1,..,m\} \quad (56)$$

$$\sum_{k=1}^{m} x_{ik} = 1, \forall\, i \in \{1,..,n\} \quad (57)$$

$$\sum_{i=1}^{n} x_{ik} = 1, \forall\, k \in \{1,..,m\} \quad (58)$$

$$x_{ik} \in \{0,1\}, \forall\, i \in \{1,..,n\}, k \in \{1,..,m\} \quad (59)$$

Where, $p_{kl}$ is a set of $\eta$ disjoint paths between substrate nodes k and l. $Length(p_{kl})$ is the total length of all substrate paths in the set $p_{kl}$. $Band(p_{kl})$ is the minimum free bandwidth in all substrate links that participate in substrate paths of $p_{kl}$.

Main goal of the objective function (54) is minimizing cost of embedding enhanced virtual networks. Cost of embedding virtual nodes depends on total CPU capacity of enhanced virtual network. Each virtual node is allocated to one and only one substrate node. Therefore, cost of embedding virtual nodes is considered unvarying in the previous formulation. In the other side, each virtual link is divided into $\eta$ links and embedded in $\eta$ substrate paths, which contain sequences of substrate links. Thus, cost of embedding virtual link depends on total lengths of all required substrate paths to accommodate this virtual link.

Constraint (55) ensures that there is sufficient substrate CPU to embed virtual node. Constraint (56) reveals that there is enough free bandwidth in all substrate links that are employed to embed virtual link.

To solve the quadratic integer program (54)-(59), the quadratic term $x_{ik} x_{jl}$ is eliminated to transform the problem into mixed integer linear program. The linearization of the problem allows us to solve the problem by applying the neural network that is proposed in [6]. After replacing quadratic term $x_{ik} x_{jl}$ with four-dimensional array $y_{ikjl}$ and replacing zero-one integrality constraints with nonnegativity constraints, the problem becomes equivalent to the following linear programming problem.

$Min\ F(Z) =$
$$\sum_{i=1}^{n}\sum_{j=1}^{n}\sum_{k=1}^{m}\sum_{l=1}^{m} Length(p_{kl}) y_{ikjl}\ b_{ij} \quad (60)$$

$$(c_i^v - c_k^s)x_{ik} \leq 0, \forall\ i \in \{1,..,n\}, k \in \{1,..,m\} \quad (61)$$

$$\left(\frac{1}{\eta-1}b_{ij} - Band(p_{kl})\right)y_{ikjl} \leq 0, \forall\ i,j \in \{1,..,n\}, k,l \in \{1,..,m\} \quad (62)$$

$$\sum_{i=1}^{n}\sum_{k=1}^{m}\sum_{j=1}^{n}\sum_{l=1}^{m} y_{ikjl} = n.m \quad (63)$$

$$2y_{ikjl} - x_{ik} - x_{jl} \leq 0, \forall\ i,j \in \{1,..,n\}, k,l \in \{1,..,m\} \quad (64)$$

$$y_{ikjl} \geq 0, \forall\ i,j \in \{1,..,n\}, k,l \in \{1,..,m\} \quad (65)$$

$$\sum_{k=1}^{m} x_{ik} = 1, \forall\ i \in \{1,..,n\} \quad (66)$$

$$\sum_{i=1}^{n} x_{ik} = 1, \forall\ k \in \{1,..,m\} \quad (67)$$

$$x_{ik} \geq 0, \forall\ i \in \{1,..,n\}, k \in \{1,..,m\} \quad (68)$$

$Z$ is defined as $(X, Y)$, where $X$ and $Y$ are vectors represent all variables in the two-dimensional array X and in the four-dimensional array Y.

By using Lagrange and Karush-Kuhn-Tucker multipliers, constraints are switched to penalties in the following Lagrangian function:

$$L(z, \lambda, \mu) = \sum_{i=1}^{n}\sum_{j=1}^{n}\sum_{k=1}^{m}\sum_{l=1}^{m} Length(p_{kl}).y_{ikjl}.b_{ij}$$

$$+ \lambda_1\left(\sum_{i=1}^{n}\sum_{k=1}^{m}\sum_{j=1}^{n}\sum_{l=1}^{m} y_{ikjl} - n.m\right)$$

$$+ \sum_{i=1}^{n} \lambda_{i+1}.\left(\sum_{k=1}^{m} x_{ik} - 1\right)$$

$$+ \sum_{k=1}^{m} \lambda_{k+(n+1)}.\left(\sum_{i=1}^{n} x_{ik} - 1\right)$$

$$+ \sum_{i=1}^{n}\sum_{k=1}^{m}\left(\mu_{k+m.(i-1)}.x_{ik}.(c_i^v - c_k^s)\right)$$

$$+ \sum_{i=1}^{n}\sum_{k=1}^{m}\sum_{j=1}^{n}\sum_{l=1}^{m}\left(\mu_{n.m+q_{ikjl}}.y_{ikjl}.\left(\frac{1}{\eta-1}b_{ij} - Band(p_{kl})\right)\right)$$

$$+ \sum_{i=1}^{n}\sum_{k=1}^{m}\sum_{j=1}^{n}\sum_{l=1}^{m}\left(\mu_{n.m+n^2.m^2+q_{ikjl}}.(2y_{ikjl} - x_{ik} - x_{jl})\right), \quad (69)$$

$$q_{ikjl} = i + n.(k-1) + n.m.(j-1) + n^2.m.(l-1)$$

Where, $\lambda = (\lambda_1,..,\lambda_i), i \in \{1,..,n+m+1\}$ is vector of Lagrange multipliers and $\mu = (\mu_1,..,\mu_i), i \in \{1,..,n.m+2n^2.m^2\}$ is vector of Karush-Kuhn-Tucker multipliers. The dual problem is formulated as follows:

Maximize

$$D(\lambda, \mu) = -n.m.\lambda_1 - \sum_{i=1}^{n} \lambda_{i+1} - \sum_{k=1}^{m} \lambda_{k+(n+1)} \quad (70)$$

Subject to

$$Length(p_{kl}).b_{ij} + \lambda_1 + \mu_{n.m+q_{ikjl}}.\left(\frac{1}{\eta-1}b_{ij} - Band(p_{kl})\right) + 2\mu_{n.m+n^2.m^2+q_{ikjl}} = 0, \forall\ i,j \in \{1,..,n\}, k,l \in \{1,..,m\}, q_{ikjl} = i + n.(k-1) + n.m.(j-1) + n^2.m.(l-1) \quad (71)$$

$$\lambda_{i+1} + \lambda_{k+(n+1)} + \mu_{k+m.(i-1)}.(c_i^v - c_k^s) - \sum_{j=1}^{n}\sum_{l=1}^{m} \mu_{n.m+n^2.m^2+q_{ikjl}} - \sum_{j=1}^{n}\sum_{l=1}^{m} \mu_{n.m+n^2.m^2+q_{jlik}} = 0, \forall\ i \in \{1,..,n\}, k \in \{1,..,m\}, q_{ikjl} = i + n.(k-1) + n.m.(j-1) + n^2.m.(l-1) \quad (72)$$

$$\mu_i \geq 0, \forall\ i \in \{1,..,n.m + 2.n^2.m^2\} \quad (73)$$

Finally, the dynamic equation of the neural network is defined by the following differential equations:

$$\frac{dy_{ikjl}}{dt} = -\beta\Big\{Length(p_{kl}).b_{ij}.\Big(\sum_{s,q=1}^{n}\sum_{r,v=1}^{m} Length(p_{rv}).y_{srqv}.b_{sq} + n.m.\lambda_1 + \sum_{i=1}^{n} \lambda_{i+1} + \sum_{k=1}^{m} \lambda_{k+(n+1)}\Big) + y_{ikjl} - |y_{ikjl}| + \sum_{s=1}^{n}\sum_{q=1}^{n}\sum_{r=1}^{m}\sum_{v=1}^{m} y_{srqv} - n.m + \Big(Band(p_{kl}) - \frac{1}{\eta-1}b_{ij}\Big)^2.y_{ikjl} - \Big(Band(p_{kl}) - \frac{1}{\eta-1}b_{ij}\Big)\Big|\Big(Band(p_{kl}) - \frac{1}{\eta-1}b_{ij}\Big).y_{ikjl}\Big| - 2x_{ik} - 2x_{jl} + 4y_{ikjl} + 2|x_{ik} + x_{jl} - 2y_{ikjl}|\Big\}, \forall\ i,j \in \{1,..,n\}, k,l \in \{1,..,m\} \quad (74)$$

$$\frac{dx_{ik}}{dt} = -\beta\{x_{ik} - |x_{ik}| + \sum_{r=1}^{m} x_{ir} + \sum_{q=1}^{n} x_{qk} - 2 + (c_k^s - c_i^v)^2.x_{ik} - (c_k^s - c_i^v).|(c_k^s - c_i^v).x_{ik}| + \sum_{j=1}^{n}\sum_{l=1}^{m}(x_{ik} + x_{jl} - 2y_{ikjl}) + \sum_{j=1}^{n}\sum_{l=1}^{m}(x_{ik} + x_{jl} - 2y_{jlik}) - 2x_{ik} + 2y_{ikik} - \sum_{j=1}^{n}\sum_{l=1}^{m}|x_{ik} + x_{jl} - 2y_{ikjl}| - \sum_{j=1}^{n}\sum_{l=1}^{m}|x_{ik} + x_{jl} - 2y_{jlik}| + |2x_{ik} + 2y_{ikik}|\}, \forall\ i \in \{1,..,n\}, k \in \{1,..,m\} \quad (75)$$

$$\frac{d\lambda_1}{dt} =$$
$$-\beta \left\{ n.m. \left( \sum_{i=1}^n \sum_{j=1}^n \sum_{k=1}^m \sum_{l=1}^m Length(p_{kl}) y_{ikjl} \, b_{ij} + n.m.\lambda_1 + \sum_{i=1}^n \lambda_{i+1} + \sum_{k=1}^m \lambda_{k+(n+1)} \right) + \right.$$
$$\sum_{i=1}^n \sum_{k=1}^m \sum_{j=1}^n \sum_{l=1}^m \left( Length(p_{kl}).b_{ij} + \lambda_1 + \mu_{n.m+q_{ikjl}} \cdot \left(\frac{1}{\eta-1} b_{ij} - Band(p_{kl})\right) + \right.$$
$$\left.\left. 2\mu_{n.m+n^2.m^2+q_{ikjl}} \right) \right\} \qquad (76)$$

$$\frac{d\lambda_{r+1}}{dt} =$$
$$-\beta \left\{ \left( \sum_{i=1}^n \sum_{j=1}^n \sum_{k=1}^m \sum_{l=1}^m Length(p_{kl}) y_{ikjl} \, b_{ij} + n.m.\lambda_1 + \sum_{i=1}^n \lambda_{i+1} + \sum_{k=1}^m \lambda_{k+(n+1)} \right) + \right.$$
$$\sum_{k=1}^m \left( \lambda_{r+1} + \lambda_{k+(n+1)} + \mu_{k+m.(r-1)} \cdot (c_r^v - c_k^s) - \right.$$
$$\left.\left. \sum_{j=1}^n \sum_{l=1}^m \left( \mu_{n.m+n^2.m^2+q_{rkjl}} + \mu_{n.m+n^2.m^2+q_{jlrk}} \right) \right) \right\}, \forall \, r \in \{1,..,n\} \qquad (77)$$

$$\frac{d\lambda_{r+(n+1)}}{dt} =$$
$$-\beta \left\{ \left( \sum_{i=1}^n \sum_{j=1}^n \sum_{k=1}^m \sum_{l=1}^m Length(p_{kl}) y_{ikjl} \, b_{ij} + n.m.\lambda_1 + \sum_{i=1}^n \lambda_{i+1} + \sum_{k=1}^m \lambda_{k+(n+1)} \right) + \right.$$
$$\sum_{i=1}^n \left( \lambda_{i+1} + \lambda_{r+(n+1)} + \mu_{r+m.(i-1)} \cdot (c_i^v - c_r^s) - \right.$$
$$\left.\left. \sum_{j=1}^n \sum_{l=1}^m \left( \mu_{n.m+n^2.m^2+q_{irjl}} + \mu_{n.m+n^2.m^2+q_{jlir}} \right) \right) \right\}, \forall \, r \in \{1,..,m\} \qquad (78)$$

$$\frac{d\mu_{k+m.(i-1)}}{dt} = -\beta \left\{ \mu_{k+m.(i-1)} - |\mu_{k+m.(i-1)}| + \right.$$
$$(c_i^v - c_k^s) \cdot \left( \lambda_{i+1} + \lambda_{k+(n+1)} + \mu_{k+m.(i-1)} \cdot (c_i^v - c_k^s) \right) -$$
$$\left.\sum_{j=1}^n \sum_{l=1}^m \left( \mu_{n.m+n^2.m^2+q_{ikjl}} + \mu_{n.m+n^2.m^2+q_{jlik}} \right) \right\}, \forall \, i \in \{1,..,n\}, k \in \{1,..,m\} \qquad (79)$$

$$\frac{d\mu_{n.m+q_{ikjl}}}{dt} = -\beta \left\{ \mu_{n.m+q_{ikjl}} - |\mu_{n.m+q_{ikjl}}| + \right.$$
$$\left( \frac{1}{\eta-1} b_{ij} - Band(p_{kl}) \right) \cdot \left( Length(p_{kl}).b_{ij} + \lambda_1 + \mu_{n.m+q_{ikjl}} \cdot \left( \frac{1}{\eta-1} b_{ij} - Band(p_{kl}) \right) + \right.$$
$$\left.\left. 2\mu_{n.m+n^2.m^2+q_{ikjl}} \right) \right\}, \forall \, i,j \in \{1,..,n\}, k,l \in \{1,..,m\} \qquad (80)$$

$$\frac{d\mu_{n.m+n^2.m^2+q_{ikjl}}}{dt} =$$
$$-\beta \left\{ \mu_{n.m+n^2.m^2+q_{ikjl}} - |\mu_{n.m+n^2.m^2+q_{ikjl}}| + \right.$$
$$2 \, Length(p_{kl}).b_{ij} + 2\lambda_1 + 2\mu_{n.m+q_{ikjl}} \cdot \left( \frac{1}{\eta-1} b_{ij} - Band(p_{kl}) \right) + 4\mu_{n.m+n^2.m^2+q_{ikjl}} - \lambda_{i+1} - \lambda_{k+(n+1)} -$$
$$\mu_{k+m.(i-1)} \cdot (c_i^v - c_k^s) +$$
$$\sum_{r=1}^n \sum_{u=1}^m \left( \mu_{n.m+n^2.m^2+q_{ikru}} + \mu_{n.m+n^2.m^2+q_{ruik}} \right) -$$
$$\lambda_{j+1} - \lambda_{l+(n+1)} - \mu_{l+m.(j-1)} \cdot (c_j^v - c_l^s) +$$
$$\sum_{r=1}^n \sum_{u=1}^m \left( \mu_{n.m+n^2.m^2+q_{jlru}} + \right.$$
$$\left.\left. \mu_{n.m+n^2.m^2+q_{rujl}} \right) \right\}, \forall \, i,j \in \{1,..,n\}, k,l \in \{1,..,m\} \qquad (81)$$

## IV. EVALUATION

To evaluate the performance of the proposed approach (CND-SVNE), its performance has been compared with Failure Independent Protection (FIP) approach [1]. FIP adds one redundant node and set of links to connect this node with all remaining nodes. Three metrics are used in the evaluation: VNE revenue, VN acceptance ratio, and substrate resources utilization. Where VNE revenue is the sum of all accepted and accommodated virtual resources, VN acceptance ratio is number of accepted virtual networks divided by total number of submitted virtual networks, and substrate resources utilization is used substrate resources divided by total substrate resources.

In the Evaluation environment, substrate network topology has been generated with 100 nodes and 500 links by using Waxman generator. Bandwidth of the substrate links are uniformly distributed between 50 and 150 with average 100. Each substrate node is randomly assigned one of the following server configurations: HP ProLiant ML110 G4 (Intel Xeon 3040, 2 cores X 1860 MHz, 4 GB), or HP ProLiant ML110 G5 (Intel Xeon 3075, 2 cores X 2660 MHz, 4 GB). We generated 1000 Virtual network topologies using Waxman generator with average connectivity 50%. The number of virtual nodes in each VN is variant from 2 to 20. Each virtual node is randomly assigned one of the following CPU: 2500 MIPS, 2000 MIPS, 1000 MIPS, and 500 MIPS, which are correspond to the CPU of Amazon EC2 instance types. Bandwidths of the virtual links are real numbers uniformly distributed between 1 and 50. VN's arrival times are generated randomly with arrival rate 10 VNs per 100 time units. The lifetimes of the VNRs are generated randomly between 300 and 700 time units with average 500 time units. Generated SN and VNs topologies are stored in brite format and used as inputs for all mapping algorithms.

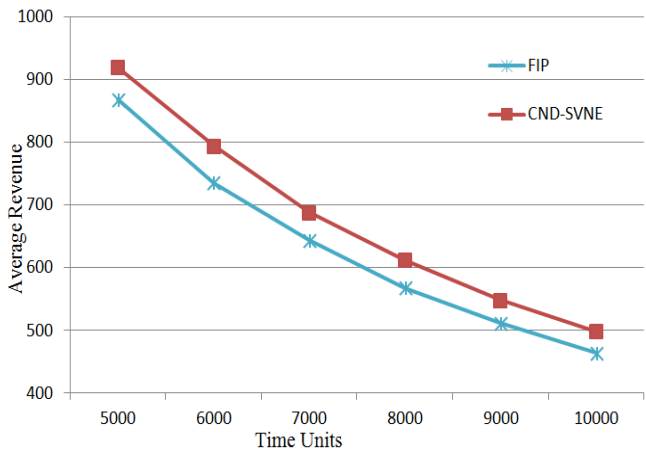

Fig. 1. Revenue comparison.

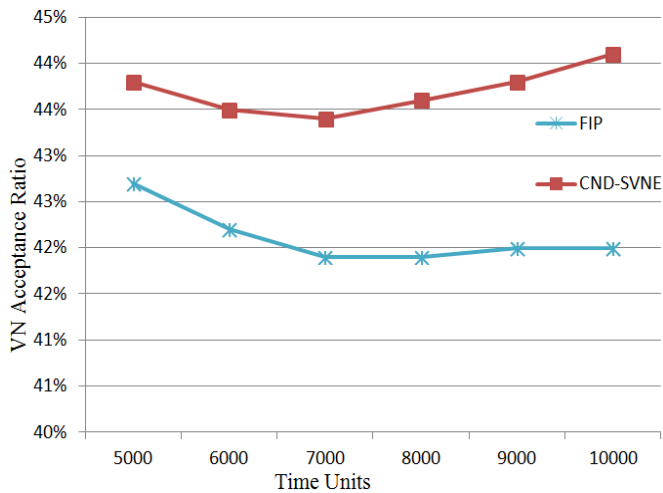

Fig. 2. VN Acceptance ratio comparison.

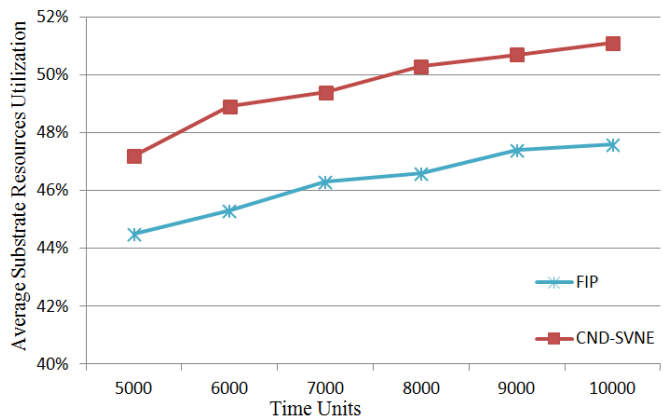

Fig. 3. Substrate resources utilization comparison.

As shown in Fig. 1, 2, and 3, the proposed approach increases VN acceptance ratio compared with FIP. By accepting and accommodating more virtual network requests, the proposed approach increases datacenter's revenue and increases substrate resources utilization. This improvement comes from reducing amount of required redundant virtual links in the proposed approach. Additionally, the proposed approach does not require any additional computation after failures this is due to existence of recovery plan (mutation matrix) for each failure.

## V. CONCLUSION

In this paper, a collective neurodynamic survivable virtual network embedding approach has been proposed. The proposed approach combines substrate node failure survivable virtual network embedding approach proposed by Guo et al. [1] with substrate link failure survivable virtual network embedding approach proposed by Khan et al. [7]. Fast convergence of the neural network optimization that has been proposed by Xia [6] has been exploited to reduce amount of required redundant resources, while co-ordination between neurodynamic models has been done by using particle swarm optimization. Experimental results show that the proposed approach outperforms Failure Independent Protection approach.

For the future work, we plan to extend the proposed approach to consider virtual network size during specifying redundant resources. Instead of adding fixed number of redundant nodes for each virtual network, number of redundant nodes will be proportional to size of virtual network. Furthermore, influence of the proposed approach on fragmentation of substrate resources will be investigated for further improvement.

ACKNOWLEDGMENT

The authors would like to express their cordial thanks to the department of Research and Development (R&D) of IMAM, university for research grant no: 370903.